\documentclass[aps,english,showpacs,twocolumn]{revtex4-1}
\usepackage{amssymb}
\usepackage{amsmath}
\usepackage{graphicx}
\usepackage{epsfig}

\begin{document}

\title{Dynamics of a tight-binding ring threaded by time-periodic magnetic
flux}
\author{W.H. Hu and Z. Song}
\email[E-mail: ]{songtc@nankai.edu.cn}
\affiliation{School of Physics, Nankai University, Tianjin 300071, China}

\begin{abstract}
We analytically study the effects of periodically alternating magnetic
fields on the dynamics of a tight-binding ring. It is shown that an
arbitrary quantum state can be frozen coherently at will by the very
frequent square-wave field as well as the monochromatic-wave field when the
corresponding optimal amplitudes are taken. Numerical simulations show that
the average fidelity depends on not only the system parameters, but also the
features of the quantum state. Moreover, taking the initial zero-momentum
Gaussian wave packets as examples, we show the dependence of the threshold
frequency on the width of the packet for the given average fidelities.
These observations provide a means to perform the quantum state engineering.
\end{abstract}

\pacs{03.67.-a, 03.75.-b, 03.75.Lm}
\maketitle

\section{Introduction}

Coherent quantum state storage and transfer via a coupled qubit system are
an important problem in the emerging area of quantum information processing
(QIP). One of the promising method of quantum state transfer is employing a
solid-state data bus with minimal spatial and dynamical control over the
on-chip interactions between qubits \cite{Bose1,Ekert,ST,LY,key-1,SZ,TJO04}.
However, small imperfections in receiving a quantum state and storing it
coherently can seriously affect the fidelity of QIP. Stopping and freezing a
flying qubit within a region of the data bus is a tool for this task. It has
been proposed that a coupled cavity array system exhibits the possibility of
an all-optical coherent control of light \cite{Fan1,Fan2,Fan3,ZhouL}. The
dynamical control includes adiabatic scheme, under which the quantum state
is fixed on the superposition of the instantaneous eigenstates when the
Hamiltonian varies slowly, and bang-bang control techniques, by means of a
dynamical control field, at averaging to cease the unwanted evolution of the
state.

Dynamical decoupling (DD) is a well established paradigm of bang-bang
control techniques, which employs a specially designed sequence of control
pulses applied to the qubits in order to negate the coupling of the central
spins to their environment \cite{DD}. Moreover, the quantum Zeno effect has
been proposed as a strategy to protect coherence \cite{DD1,DD2}. Recently,
it is proposed that a periodically driven potential can suppress the
tunneling between adjacent sites in a lattice \cite{OPL1,OPL2}. In this
paper we will pay attention to a fundamental aspect of QIP and generally
study the influence of periodically alternating magnetic fields on the
dynamics of a quantum state on a tight-binding ring. We consider the
dynamics of the states in a tight-binding ring system that is pierced by a
time-periodic magnetic flux $\Phi \left(t\right)=\Phi _{0}+\Phi _{\mathrm{A}%
}f\left( \omega t\right) $ with the angular frequency $\omega $. We
investigate the impact that the amplitude and frequency of the flux might
have on the efficacy of the quantum control. We focus on the suppression of
the evolution through a bang-bang control procedure, and study how the
occurrence of a controlling field modifies the effectiveness of the control
procedure. Our analysis is focused on the behavior of the fidelity of the
evolved state with respect to the initial state, which has been employed to
measure the efficiency of quantum state transfer. We will show that the time
evolution of a state in such a time-dependent Hamiltonian can then be
treated as an adiabatic process without any approximation. Then analytical
results can be obtained, which should give more insight into quantum
measurement and control. Additionally, this scheme can be applied to a
neutral-particle system by torsional oscillation.

In Sec.~\ref{sec:formalism} we derive a general formalism for such a
time-dependent system. We further introduce the expressions for the
time-averaged fidelity, thus completing the description of the
controllability of the systems for the quantum states. Sec.~\ref{sec:SquWav}%
,~\ref{sec:MonWav} are devoted to the applications of the formalism. These
include a detailed treatment and computation of the time evolutions of
typical initial states under the square and monochromatic time-periodic
flux, respectively. Final conclusions and discussions are drawn in Sec.~\ref%
{sec:Dis}.

\section{Model and general formalism}

\label{sec:formalism}

In this section, we present the charged particle model under consideration,
a simple tight-binding model in an external magnetic field. Here, the
particle-particle interaction is ignored for simplicity. Our approach is
based on our previous work in Ref.~\cite{YS1}, where we have proposed a
scheme for quantum state transfer. It employed a loop enclosing a static
magnetic flux to control the speed of a wave packet. Another basic operation
for the quantum-state engineering is coherently freezing a state on demand.
For instance, quantum information processing requires transferring, stopping
and freezing a flying qubit within a region of the data bus. In Ref.~\cite%
{YS1}, we studied how to move a Gaussian wave packet at a certain speed on
demand by a static magnetic flux. In this work we aim at employing the same
system with a periodically alternating flux for freezing a wave packet.
Combining the two schemes, one can accomplish the task of ``coherently
storage-transfer-storage'' of a quantum state. Now we will generalize this
description of the system in Ref.~\cite{YS1}\ by allowing for an additional
time-periodic flux. We restrict our attention to the influence of the
applied periodically alternating field on the dynamics of the particles.

\begin{figure}[tbp]
\includegraphics[ bb=0 195 480 782,width=0.45\textwidth,clip]{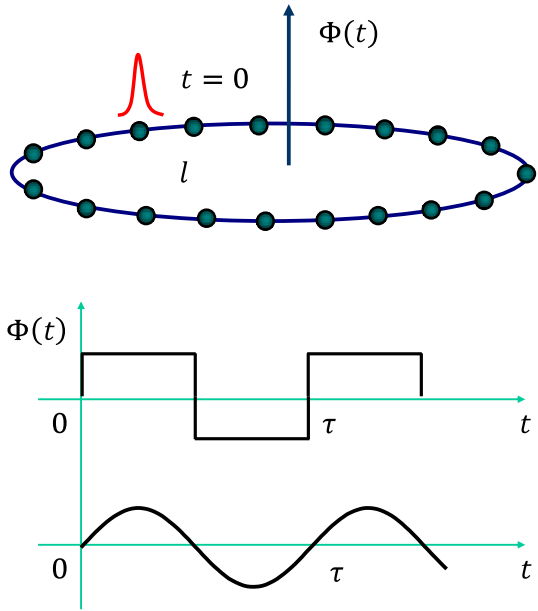}
\caption{(Color online) The schematic illustration for a tight-binding ring
threaded by a time-periodic magnetic field. The time-dependent flux for
demonstrating the influence on the evolution of a quantum state is in two
types: a square and a monochromatic waves, respectively, with amplitude
$\phi _{\mathrm{A}}$ and period $\tau$.}
\label{illus}
\end{figure}

Consider a ring lattice with $N$ sites threaded by a magnetic field
illustrated schematically in Fig.~\ref{illus}. The Hamiltonian of the
corresponding tight-binding model
\begin{equation}
H\left(t\right)=-J\sum_{j=1}^{N}\left(e^{i2 \pi \Phi \left(t\right)
/N}a_{j}^{\dagger }a_{j+1}+\mathrm{H.c.}\right)
\label{H_phi}
\end{equation}
depends on the magnetic flux through the ring in units of the flux quantum
$\Phi_{0}=h/e$. Here $a_{j}^{\dagger}$ is the creation operator of a
particle at the $j$th site with the periodic boundary condition $a_{N+1}=a_%
{1}$. The flux does not exert force on the Bloch electron, but can change
the local phase of its wave function due to the Aharonov-Bohm (AB) effect.
Note that the particle is not restricted to be either fermion or boson.

Taking the transformation
\begin{equation}
a_{j}=\frac{1}{\sqrt{N}}\sum_{k}e^{ikj}a_{k},
\label{a_j}
\end{equation}
where $k=2\pi n/N$, $n\in \left[1,N\right] $, the Hamiltonian can be readily
written as
\begin{equation}
H=-2J\sum_{k}\cos \left( k+\phi \left(t\right) \right) a_{k}^{\dagger}a_{k},
\label{H_k}
\end{equation}
with $\phi \left(t\right)=2\pi \Phi \left(t\right)/N$ and the corresponding
eigenstates in the form of
\begin{equation}
\left\vert k\right\rangle =\frac{1}{\sqrt{N}}\sum_{j}e^{ikj}\left\vert
j\right\rangle .
\label{a_k}
\end{equation}
Note that the time-dependent Hamiltonian possesses fixed eigenstates, while
the flux solely affects the eigenvalues. This will be crucial to employ such
kind of setup to investigate the control of a quantum state due to the
rareness of the exact solutions to a time-dependent Hamiltonian. The
evolution of an arbitrary state under the Hamiltonian $H$ is dictated by the
unitary operator $U\left( t^{\prime },t\right) =\exp \left( -i{%
\int_{t}^{t^{\prime }}}H\mathrm{d}t^{\prime \prime }\right) $, which yields
the propagator represented in the momentum and spatial eigenstates as
\begin{eqnarray}
U_{k^{\prime }k}\left( t^{\prime },t\right) &=&\left\langle k^{\prime
}\right\vert U\left( t^{\prime },t\right) \left\vert k\right\rangle =e^{i2J
f_{k}\left( t^{\prime },t\right) }\delta _{kk^{\prime }},  \notag \\
U_{j^{\prime }j}\left( t^{\prime },t\right) &=&\left\langle j^{\prime
}\right\vert U\left( t^{\prime },t\right) \left\vert j\right\rangle
\label{U_kk} \\
&=&\frac{1}{N}\sum_{k}e^{ik\left( j^{\prime }-j\right) }e^{i2Jf_{k}\left(
t^{\prime },t\right) },  \notag
\end{eqnarray}
where
\begin{equation}
f_{k}\left( t^{\prime },t\right) =\int_{t}^{t^{\prime }}\cos \left[ k+\phi
\left( t^{\prime \prime }\right) \right] \mathrm{d}t^{\prime \prime }.
\label{f_tt}
\end{equation}%
We note that the propagator is in diagonal form in $k$ space.

In general, one employs the fidelity
\begin{equation}
F\left(t\right)=\left\vert \left\langle \psi \left( 0\right) \right\vert
U\left( t,0\right) \left\vert \psi \left( 0\right) \right\rangle \right\vert
\end{equation}
to characterize the relation between the target state and the evolved state
at time $t$. However, when $\phi \left( t\right) $ is a periodic function,
$F\left(t\right)$ should be oscillating. Thus the long time average will be
appropriate as a measure for the deviation from the original state.

The average fidelity is defined as%
\begin{eqnarray}
\overline{F} &=&\lim_{T\rightarrow \infty }\frac{1}{T}\int_{0}^{T}F\left(
t\right) \mathrm{d}t
\label{F_ave} \\
&=&\lim_{T\rightarrow \infty }\frac{1}{T}\int_{0}^{T}\left\vert
\sum_{k}\left\vert c_{k}\right\vert ^{2}e^{i2Jf_{k}\left( t\right)
}\right\vert \mathrm{d}t,  \notag
\end{eqnarray}%
where
\begin{equation*}
c_{k}=\left\langle k\right. \left\vert \psi \left( 0\right) \right\rangle .
\end{equation*}%
In this paper, we focus on the case of periodic $\phi \left( t\right) $ with
a period of $\tau $
\begin{equation}
\phi \left( t\right) =\phi \left( t+\tau \right) .
\end{equation}%
In the following sections, we will apply the general formalism to the cases
of square and monochromatic waves due to the following reasons: The
square-wave case is an demonstrative example since it is the simplest model
to calculate. The exact solution for this particular case is helpful to
clearly present the main idea of the scheme without involving much more
complicated calculation. And the monochromatic-wave case is a more practical
situation. We will argue that the quantum state freezing can be achieved by
the high frequency alternating flux when its amplitudes $\Phi _{\mathrm{A}}$
are optimal.

\section{Square wave}
\label{sec:SquWav}

Let us begin the discussion with the simplest case: the flux is in the form
of
\begin{equation}
\phi \left( t\right) =\phi _{0}+\phi _{\mathrm{A}}\mathrm{sgn}\left( \sin
\left( \omega t\right) \right) ,
\end{equation}
where $\omega =2\pi /\tau $ is the angular frequency and sgn indicates the
sign function. It is a toy model which demonstrates how, in principle, the
periodic alternating flux can prevent a quantum state from spreading. The
Hamiltonian can be diagonalized as
\begin{eqnarray}
H &=&-2J\sum_{k}\varepsilon _{k}\left( t\right) a_{k}^{\dagger }a_{k}, \\
\varepsilon _{k}\left( t\right) &=&\cos \left[ k+\phi _{\mathrm{A}}\mathrm{%
sgn}\left( \sin \left( \omega t\right) \right) \right]
\end{eqnarray}
where we absorbed $\phi _{0}$\ into $k$ by $k\rightarrow k-\phi _{0}$ for
the sake of simplicity. Then we have
\begin{equation}
\overline{F}\left( \tau \right) =\lim_{T\rightarrow \infty }\frac{1}{T}%
\int_{0}^{T}\left\vert \sum_{k}\left\vert c_{k}\right\vert ^{2}\exp \left(
i2J\int_{0}^{t}\varepsilon _{k}\left( t^{\prime }\right) \mathrm{d}%
t^{\prime} \right) \right\vert \mathrm{d}t.  \label{F_square1}
\end{equation}

Obviously, it is hard to get the analytical expression of $\overline{F}%
\left( \tau \right) $. However, one can get insight into the influence of
the square wave on a quantum state from the following analysis. During each
interval, with the flux being static, the dynamics of the quantum states in
such situation has been discussed in Ref.~\cite{YS1}. In particular, for $%
\phi _{\mathrm{A}}=\pi /2$ , the corresponding Hamiltonians are
\begin{equation}
H(t)=\left\{
\begin{array}{cc}
H_{+}, & \mathrm{sgn}\left( \sin \left( \omega t\right) \right) \succ 0 \\
H_{-}, & \mathrm{sgn}\left( \sin \left( \omega t\right) \right) \prec 0%
\end{array}
\right.  \label{H_t}
\end{equation}
with
\begin{equation}
H_{\pm }=\pm 2J\sum_{k}\sin ka_{k}^{\dagger }a_{k}.
\end{equation}
Then the dynamics on the successive time intervals $\left[ t,t+\tau/2\right]
$ and $\left[ t+\tau/2,t+\tau \right]$ are time reversal processes to each
other, i.e.,
\begin{eqnarray}
U_{k^{\prime}k}\left(t+\tau/2,t\right)
&=&U_{k^{\prime}k}\left(t+\tau,t\right)
U_{k^{\prime}k}\left(t+\tau/2,t+\tau\right)  \notag \\
&=&U_{k^{\prime}k}^{-1}\left(t+\tau,t+\tau/2\right) , \\
t &\in &\left[ 0,\tau /2\right] ,  \notag
\end{eqnarray}
which leads to
\begin{equation}
\left\vert \psi \left( t\right) \right\rangle =\left\vert \psi \left( t+\tau
\right) \right\rangle  \label{Periodic WF}
\end{equation}
for an arbitrary state. Then after a period of time $\tau$, any state will
go back to its initial state, i.e.,
\begin{equation}
F(t)=F(t+\tau )\text{, }F(n\tau )=1.  \label{Periodic F}
\end{equation}
For small $\tau $, the evolved state should not leave its initial state so
far at any time. Intuitively, in the limit of $\tau \rightarrow 0$, the
initial state may be frozen at its initial position. Actually, the
periodicity of $\left\vert \psi \left( t\right) \right\rangle $ admits
\begin{equation}
\overline{F}=\frac{2}{\tau }\int_{0}^{\tau /2}\left\vert \sum_{k}\left\vert
c_{k}\right\vert ^{2}e^{i2J\sin kt}\right\vert \mathrm{d}t ,
\end{equation}
and
\begin{equation}
\lim_{\tau \rightarrow 0}\overline{F}\simeq \lim_{\tau \rightarrow 0}\frac{2%
} {\tau }\int_{0}^{\tau /2}\left\vert 1+i2Jt\sum_{k}\left\vert
c_{k}\right\vert ^{2}\sin k\right\vert \mathrm{d}t =1,
\end{equation}
due to $\sum_{k}\left\vert c_{k}\right\vert ^{2}=1$, $\left\vert
\sum_{k}\left\vert c_{k}\right\vert ^{2}\sin k\right\vert \prec 1$. We
conclude that an arbitrary state can be frozen at will when the frequency is
sufficient high. Particularly, in the case of $\left\vert c_{k}\right\vert
^{2}$\ being symmetrical about $k=0$, we simply have
\begin{equation}
\overline{F}=\frac{2}{\tau }\int_{0}^{\tau /2}\left\vert \sum_{k}\left\vert
c_{k}\right\vert ^{2}\cos \left( 2Jt\sin k\right) \right\vert \mathrm{d}t,
\end{equation}
and for small $\tau $ $\left( \tau \ll J^{-1}\right) $, it has the form of
\begin{equation}
\overline{F}\simeq 1-\frac{1}{6}J^{2}\tau ^{2}\sum_{k}\left\vert
c_{k}\right\vert ^{2}\sin ^{2}k.
\end{equation}

Obviously, $\sum_{k}\left\vert c_{k}\right\vert ^{2}\sin ^{2}k$ determines
the speed of convergence: the more the contribution of $\left\vert%
c_{k}\right\vert ^{2}$ to zero $k$, the faster the convergence speed. In the
following, we estimate the convergence speed of a specific state. A Gaussian
wave packet with central momentum $k_{0}$ can be expressed in the form of
\begin{equation}
c_{k}=\lambda \exp \left( -\frac{\alpha ^{2}}{2}\left( k-k_{0}\right)
^{2}\right) ,
\label{Gaussian}
\end{equation}%
where
\begin{equation*}
\lambda ^{-1}=\sqrt{\sum_{k}\exp \left( -\alpha ^{2}\left(%
k-k_{0}\right) ^{2}\right) }
\end{equation*}
is the normalization factor and equal to $\sqrt{N/\left( 2\sqrt{\pi}\alpha%
 ^{2}\right)}$ for large $N$. Here $\alpha$ determines the width of the
wave packet. We consider such a state with $k_{0}=0$ as the initial state.
For large frequency $\nu =1/\tau $, we have the frequency dependence of the
average fidelity as
\begin{equation}
\overline{F}-1\simeq -\frac{J^{2}}{12\nu ^{2}}\left[ 1-e^{-\frac{1}{\alpha
^{2}}}\right] .
\end{equation}%
Then for the given initial state, we have evaluated the threshold frequency
as
\begin{equation}
\nu _{c}=J\sqrt{\frac{1-e^{-\alpha ^{-2}}}{12\left( 1-\overline{F_{c}}%
\right) }},
\label{V_c}
\end{equation}%
from which one can obtain the control of the quantum state with fidelity $%
\overline{F_{c}}\left( \tau \right) $. We plot Eq.~(\ref{V_c}) in Fig.~\ref%
{SquNuAlp}, which shows that the threshold frequency increases rapidly when
the width of the state gets narrow.

So far, our investigation is carried out analytically for a particular case
of $\phi _{\mathrm{A}}=\pi /2$. It can be seen that the mechanism of the
perfect quantum state freezing expressed as Eq.~(\ref{Periodic F}) is the
periodicity of the evolved wave function Eq.~(\ref{Periodic WF}) arising
from the amplitude $\phi _{\mathrm{A}}=\pi /2$. It is worthy to investigate
what happens for other values of $\phi _{\mathrm{A}}$ and finite
frequencies. In the following, we will perform a series of numerical
simulations from several aspects.

\begin{figure}[tbp]
\includegraphics[bb=0 0 595 842,width=0.45\textwidth]{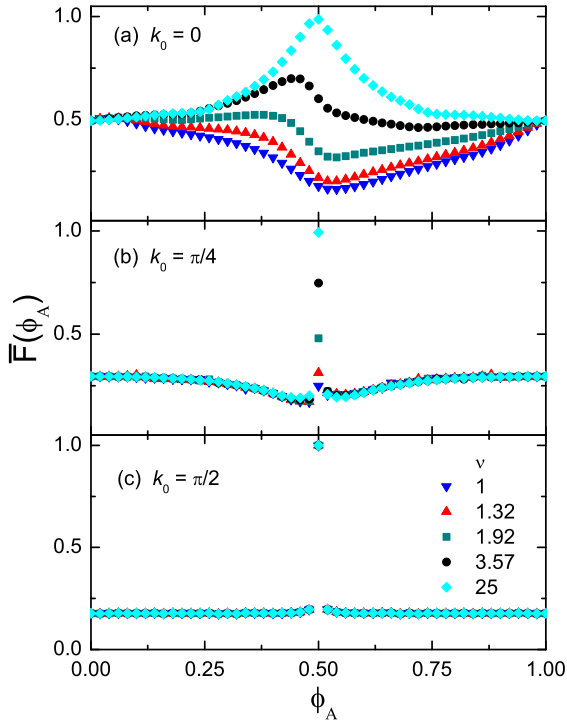}
\caption{(Color online) Average fidelity as a function of field amplitude
$\phi _{\mathrm{A}}$ (units of $\pi$) and frequency $\nu$ (units of $J$)
of a square-wave flux for the Gaussian wave packet with $\alpha =50$ and
central momenta $k_{0}=0$, $\pi /4$, $\pi /2$ on an $N=1000$ ring. The
average fidelity is computed over the time interval $\left[ 0, 25N/J\right]%
$. It shows that $\pi /2$ is the optimal amplitude and the fidelity becomes
very sensitive to the magnitude for the initial wave packet with high speed.}
\label{SquSens}
\end{figure}

First of all, we investigate the influence of the amplitude. The numerical
simulations are performed in Fig.~\ref{SquSens} for a Gaussian wave packet
which has the form of Eq.~(\ref{Gaussian}) with different central momenta in
the systems with different amplitudes and different frequencies. It shows
that the average fidelity approaches to unit for $\phi _{\mathrm{A}}=\pi /2$
when $\nu $ is sufficiently high, which is in agreement with the above
analysis. And it is noted that the locations of the maxima of the average
fidelity for moderate $\nu$ shift to the left for the case of $k_{0}=0$. It
is important for practical implementation. In the case of not sufficiently
high frequency, the optimal amplitude should be smaller than $\pi/2$. For $%
k_{0}=\pi/4$, $\pi/2$ the fidelity becomes very sensitive to the magnitude
of $\phi _{\mathrm{A}}$. This feature can be exploited to select the wave
packet with preferable $k_{0}$ in the following way. Suppose a many-particle
initial state, which consists of wave packets with various speeds. One can
first tune $\phi _{0}$ to meet $\phi_{0}+k_{0}=\pi/2$. Then take $\phi _{%
\mathrm{A}}=\pi /2$ to hold the wave packets with $k_{0}$ on demand.

\begin{figure}[tbp]
\includegraphics[bb=0 0 596 842,width=0.45\textwidth]{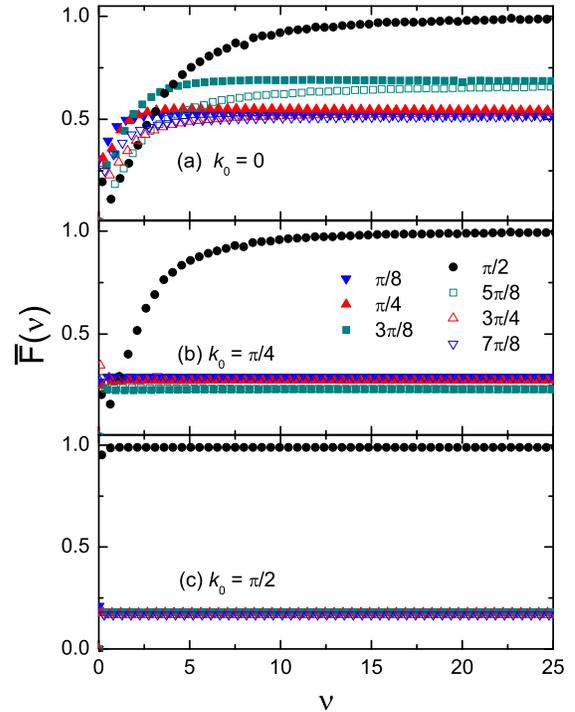}
\caption{(Color online) Average fidelity as a function of frequency of the
square-wave flux for a Gaussian wave packet with $\alpha =50$ on an
$N=1000$ ring. Panels stand for different values of the amplitudes $\phi%
_{\mathrm{A}}=\pi/8$, $\pi/4$, $3\pi/8$, $\pi/2$, $5\pi/8$, $3\pi/4$,
$7\pi/8$ and central momenta $k_{0}=0$, $\pi/4$, $\pi/2$ of the initial
wave packet. It shows that, when $\phi _{\mathrm{A}}=\pi/2$, frequently
alternating flux suppresses the evolution of the quantum
states. Here the frequency $\nu$ is expressed in units of $J$.}
\label{SquGau}
\end{figure}

\begin{figure}[tbp]
\includegraphics[bb=0 0 576 408, width=0.45\textwidth]{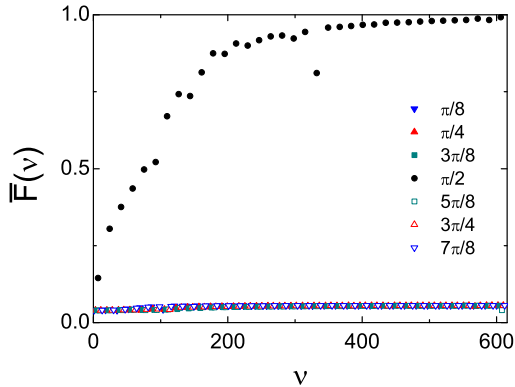}
\caption{(Color online) Average fidelity as a function of frequency
of the square-wave flux for a single-site state on an $N=1000$ ring.
Plots are presented for different values of the amplitudes $\phi_{%
\mathrm{A}}=\pi /8$, $\pi /4$, $3\pi /8$, $\pi /2$, $5\pi /8$,
$3\pi/4$ and $7\pi/8$. It shows that, when $\phi _{\mathrm{A}}=\pi/2$,
frequently alternating flux suppresses the evolution of the
quantum states, while other cases exhibit small fidelity. Here the
frequency $\nu$ is expressed in units of $J$.}
\label{SquDelta}
\end{figure}

Secondly, the analysis above is based on the assumption that not only the
amplitude of the flux is $\pi /2$, but also the frequency is sufficient
high. Then even the amplitude is taken exactly as $\pi/2$, the efficiency of
the scheme is different for different quantum states under the finite
frequency. For instance, the average fidelity during the period $\tau$
mainly depends on the overlap of the initial wave packet and its evolution
driven by $H_{\pm }$. In order to demonstrate these analysis, the numerical
simulations are performed for two kinds of initial states: a Gaussian wave
packet with central momentum $k_{0}$, which has the form of Eq.~(\ref%
{Gaussian}), and single-site state $\left\vert l\right\rangle \equiv
a_{l}^{\dagger }\left\vert 0\right\rangle $. We consider the time evolutions
of this GWP with different $k_{0}$ in the system with different amplitudes
and different frequencies. The average fidelity $\overline{F}\left(
\tau\right)$ over the interval $T\preceq 25N/J$ is plotted in Fig.~\ref%
{SquGau}. It shows the following features: (i) The average fidelity
approaches to unit for all the given initial wave packets with different $%
k_{0}$ when the amplitude is $\pi/2$, which is in agreement with the above
analysis. (ii) The threshold frequency in the case of $\phi _{\mathrm{A}%
}=\pi/2$ gets lower as $k_{0}$ goes closer to $\pi/2$. This also accords
with the above analysis. Actually, the velocity of a $\pi/2$ wave packet
becomes zero under the Hamiltonians $H_{\pm}$. Thus it deviates from the
initial state slightly during the interval $\tau $, leading to a high
fidelity. (iii) The optimal average fidelity becomes more sensitive as
$k_{0}$ goes closer to $\pi/2$, which is in agreement with the results in
Fig.~\ref{SquSens}.

Based on these features, one can design a scheme to achieve the maximal
fidelity. In ideal case, for any given frequency of the field, $\phi_{%
\mathrm{A}}=\pi /2$ is always preferable. When the frequency is not
sufficient high, one can tune $\phi _{0}$ to match $k_{0}$ in order to
achieve a lower threshold frequency. However, the accuracy of the field may
affect the fidelity due to the sensitivity of it around $\phi _{\mathrm{A}%
}=\pi /2$ in practice. Then one can tune $\phi_{0}$ to stabilize the
fidelity.

\begin{figure}[tbp]
\includegraphics[bb=0 0 576 408, width=0.45\textwidth]{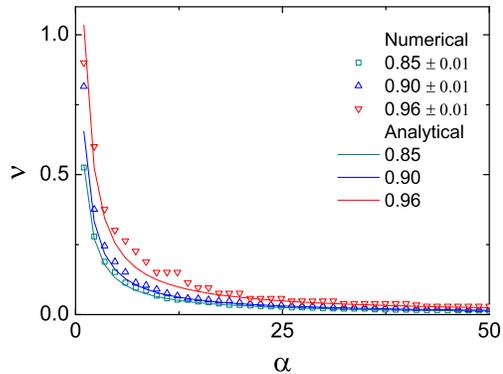}
\caption{(Color online) Threshold frequency on the widths of the
zero-momentum Gaussian wave packet. The plots are Eq.~(\ref{V_c}) and
numerical results corresponding to the average fidelities as $0.96\pm 0.01$,
$0.90\pm 0.01$ and $0.85\pm 0.01$. It shows that the threshold frequency
increases rapidly when the width of the state gets narrow.}
\label{SquNuAlp}
\end{figure}

Finally, in order to demonstrate the applicability of our findings to
control a quantum state, we also plot the average fidelity for a single-site
state $\left\vert l\right\rangle \equiv a_{l}^{\dagger }\left\vert
0\right\rangle $ in Fig.~\ref{SquDelta}. Such a state has $\left\vert
c_{k}\right\vert ^{2}=1/N$ and is the narrowest limit of a wave packet. It
shows that the average fidelity approaches to unit only in the case of $\phi
_{\mathrm{A}}=\pi /2$ and a relative high frequency. And to demonstrate the
efficiency of this method, we take a initial zero-momentum Gaussian wave
packet as examples. Numerical simulation is performed in Fig.~\ref{SquNuAlp}
for the dependence of the threshold frequency on the widths for the given
average fidelities. For a comparison we draw the curves from Eq.~(\ref{V_c})
and numerical results corresponding to the average fidelities as $0.96\pm
0.01$, $0.90\pm 0.01$ and $0.85\pm 0.01$. The width coefficient $\alpha$ of
the wave packet is taken from $1$ to $50$, ranging from a single-site state
to a very wide wave packet which is approximately a plane wave with zero
momentum. As can be seen from the figure, analytical and numerical results
both indicate that the threshold frequency increases rapidly when the width
of the state gets narrow. Then for a finite-frequency field, such a scheme
has a high efficiency for a broad wave packet.

\section{Monochromatic wave}
\label{sec:MonWav}

The simplicity of the square-wave field makes it easy to take an analytical
investigation for the concerned problems, because the exact solution of this
model is helpful to clearly present the main idea of the scheme without
involving much more complicated calculation. However, such a toy model is
not exactly accessible in experiments due to the sudden change of the flux.
In this section, we will consider the monochromatic-wave field, which is
more practical. It will been shown analytically and numerically that both
cases are similar qualitatively.

\begin{figure}[htb]
\includegraphics[bb=0 0 595 842,width=0.45\textwidth]{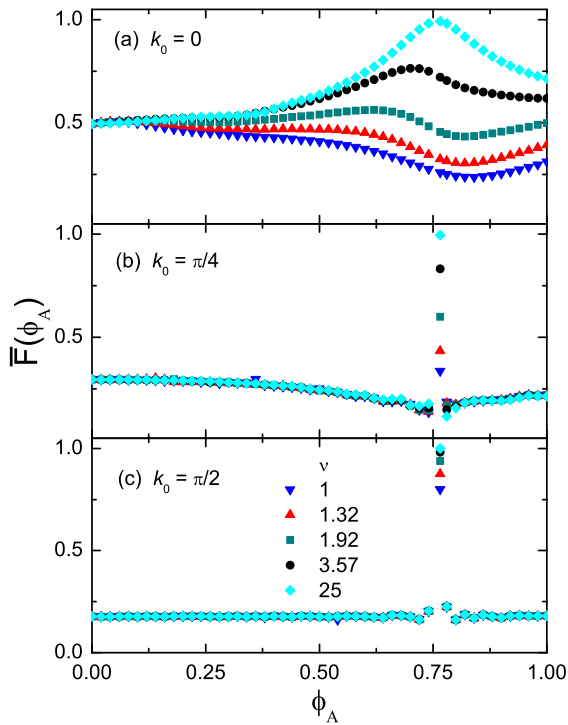}
\caption{(Color online) Average fidelity as a function of field amplitude
$\phi _{\mathrm{A}}$ (units of $\pi$) and frequency $\nu$ (units of $J$) of
a monochromatic-wave flux for the Gaussian wave packet with $\alpha =50$
and central momenta $k_{0}=0$, $\pi/4$, $\pi/2$, on an $N=1000$ ring. The
average fidelity is computed over the time interval $\left[ 0,25N/J\right]$.
It shows that $0.765\pi $ is the optimal amplitude and the fidelity
becomes very sensitive to the magnitude for the initial wave packet with high
speed.} \label{SinSens}
\end{figure}

The monochromatic-wave field is in the form of
\begin{equation}
\phi \left( t\right) =\phi _{0}+\phi _{\mathrm{A}}\sin \omega t.
\label{sine}
\end{equation}%
Unlike the square-wave, even in the special case of $\phi _{\mathrm{A}}=\pi
/2$, the fidelity is no longer a periodic function due to the breaking of
the time reversal symmetry. Thus one should consider the integral to the
whole time duration. However, the analytical function of $\phi \left(
t\right) $\ may lead to some analytical results. Here we still neglect $\phi
_{0}$ for simplicity. We firstly investigate some special cases analytically
to seek the optimal $\phi _{\mathrm{A}}$ satisfying the relation Eq.~(\ref%
{Periodic F}), and then perform numerical simulations for more general cases.

\begin{figure}[tbp]
\includegraphics[bb=0 0 596 842, width=0.45\textwidth]{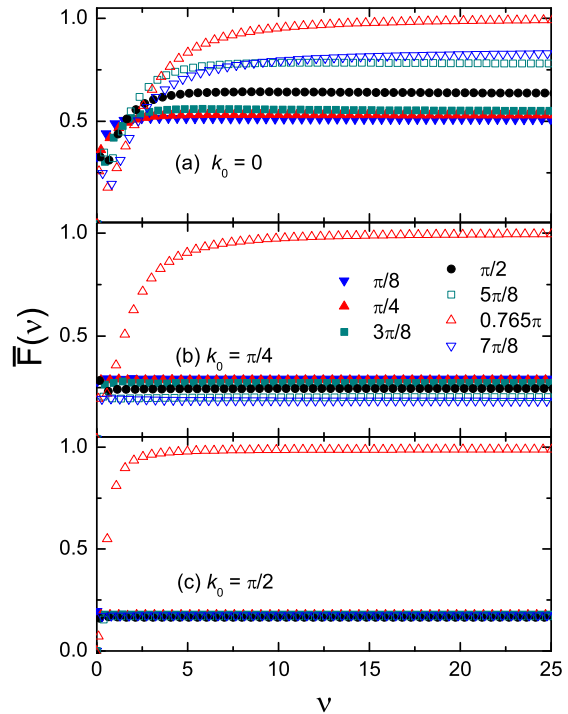}
\caption{(Color online) Average fidelity as a function of frequency of the
monochromatic-wave flux for a Gaussian wave packet with $\alpha=50$ on
an $N=1000$ ring. Panels stand for different values of the amplitudes
$\phi _{\mathrm{A}}=\pi /8$, $\pi /4$, $3\pi /8$, $\pi /2$, $5\pi /8$,
$0.765\pi$, $7\pi /8$ and central momenta $k_{0}=0$, $\pi /4$, $\pi /2$
of the initial wave packet. It shows that, when $\phi _{\mathrm{A}}=0.765%
\pi$, frequently alternating flux suppresses the evolution of the quantum
states. Here the frequency $\nu$ is expressed in units of $J$.}
\label{SinGau}
\end{figure}

\begin{figure}[tbp]
\includegraphics[bb=0 0 576 408, width=0.45\textwidth]{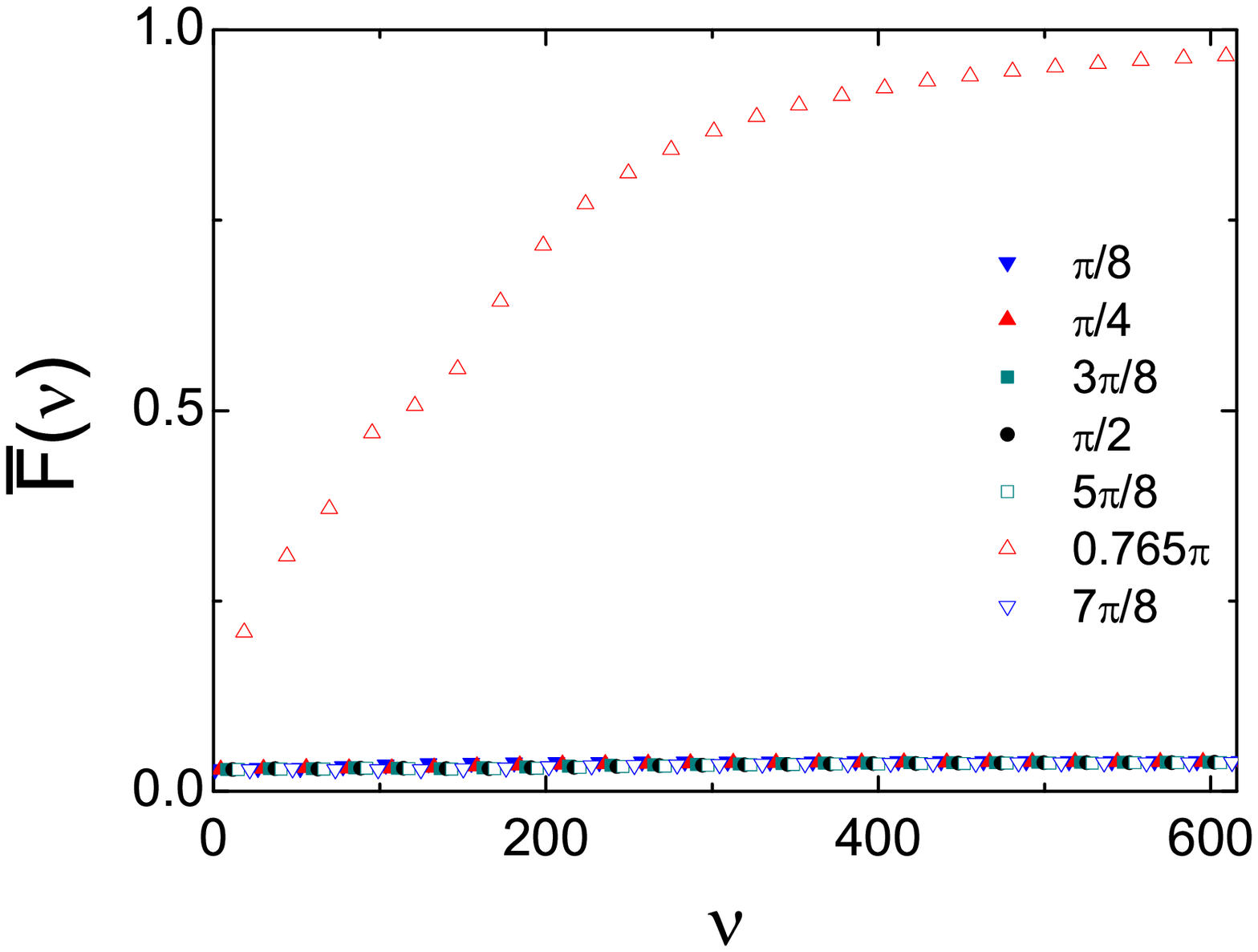}
\caption{(Color online) Average fidelity as a function of frequency of the
monochromatic-wave flux for a single-site state on an $N=1000$ ring. Plots
are presented for different values of the amplitudes $\phi _{\mathrm{A}}=%
\pi/8$, $\pi/4$, $3\pi/8$, $\pi/2$, $5\pi/8$, $0.765\pi$ and $7\pi/8$. It
shows that, when $\phi _{\mathrm{A}}= 0.765\pi$, frequently alternating
flux suppresses the evolution of the quantum states, while other cases
exhibit small fidelity. Here the frequency $\nu$ is expressed in units of
$J$.}
\label{SinDelta}
\end{figure}

\begin{figure}[tbp]
\includegraphics[bb=0 0 576 408,width=0.45\textwidth]{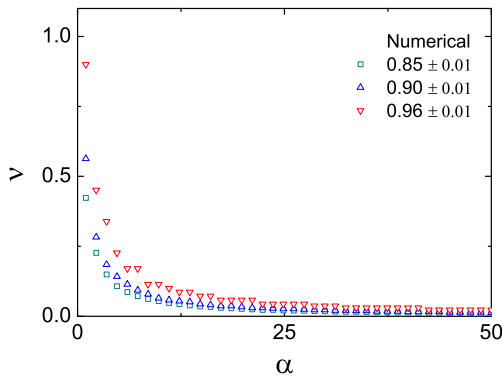}
\caption{(Color online) The same as Fig.~\ref{SquNuAlp}, but only numerical
result for the monochromatic-wave field.}
\label{SinNuAlp}
\end{figure}

Considering the evolved state at the instant $n\tau $, where $n$ is an
integer, we have
\begin{eqnarray}
f_{k}\left( n\tau \right) &=&\int_{0}^{n\tau }\cos \left[ k+\phi _{\mathrm{A}%
}\sin \omega t\right] \mathrm{d}t^{\prime }  \label{f_kn} \\
&=&n\tau \cos k\mathcal{J}_{0}\left( \phi _{\mathrm{A}}\right) .  \notag
\end{eqnarray}%
Here
\begin{equation}
\mathcal{J}_{m}\left( x\right) =\frac{1}{\pi }\int_{0}^{\pi }\cos \left(
m\theta -x\sin \left( \theta \right) \right) \mathrm{d}\theta ,
\label{Bessiel}
\end{equation}%
are the Bessel functions of the first kind. Then the corresponding fidelity
is
\begin{equation}
F\left( n\tau \right) =\left\vert \sum_{k}\left\vert c_{k}\right\vert
^{2}e^{i2Jn\tau \cos k \mathcal{J}_{0}\left( \phi _{\mathrm{A}}\right)
}\right\vert .  \label{F_nt}
\end{equation}
Note that taking $\phi _{\mathrm{A}}=0.765\pi $, we have $\mathcal{J}%
_{0}\left( \phi _{\mathrm{A}}\right) =0$, which leads to $F\left( n\tau
\right) =1$ for an arbitrary initial state. This fact is quite similar to
the case of the square-wave field with $\phi _{\mathrm{A}}=\pi /2$.

Now we turn on our numerical investigation to the more general cases. We
perform the numerical simulations for the same states discussed in the last
section. The numerical results are plotted in Figs.~\ref{SinSens},~\ref%
{SinGau} and~\ref{SinDelta}. They show that the square and monochromatic
waves lead to the similar result but with different optimal $\phi _{\mathrm{%
A}}$. The corresponding numerical result for the threshold frequency $\nu
_{c}\left( \alpha \right) $ as a function of the width of the wave packet is
plotted in Fig.~\ref{SinNuAlp}.

Based on the numerical results presented in the two above sections, we
conclude that the evolution of a quantum state can be suppressed through the
time-periodic flux. In both situations, the efficiency of the schemes
depends on the parameters $\phi _{\mathrm{A}}$, $\phi _{0}$, $\omega$, $k_{0}
$ and $\alpha$ in the similar manner. The features can be exploited to
control quantum dynamics for quantum information and computation purposes.

\section{Discussion}
\label{sec:Dis}

We have seen that the threaded magnetic flux, in stead of the electric
field, plays an important role in controlling a state. It can be applied to
a more extended system to control an uncharged particle. Actually, if the
system is rotated, an effective magnetic field will be induced in the
rotating frame of references. Therefore, a neutral particle state in a ring
lattice can be controlled via torsional oscillation.

For a rotating ring with angular frequency $\Omega $, an additional term
\begin{equation}
H_{R}=-\Omega L_{z}=-\Omega K\sum_{j=1}^{N}\left( ia_{j}^{\dagger }a_{j+1}+%
\mathrm{H.c.}\right)
\end{equation}
should be added on the Hamiltonian with $\phi =0$ in the non-inertial frame
\cite{Bhat06}, where $K$ is a constant depends on the geometry of the ring.

In summary, we have studied the influence of periodically alternating
magnetic fields on the dynamics of a quantum state on a tight-binding ring.
Our analytical and numerical calculations indicate that the evolution of a
quantum state can be suppressed through the time-periodic flux. The
efficiency of the scheme depends on not only the system parameters $\phi _{%
\mathrm{A}}$, $\phi _{0}$ and $\omega $, but also the state parameters $k_{0}
$ and $\alpha $. Based on the features of the dynamics, one can choose an
appropriate system to freeze a given state with an expected average
fidelity. It can be also exploited to select and hold a specific wave packet
among the many-body particles, thus providing a means to perform the quantum
state engineering. We expect that such an observation has applications for
information processing and quantum device physics.

\acknowledgments We acknowledge the support of the CNSF (Grant No. 10874091)
and National Basic Research Program (973 Program) of China under Grant No.
2012CB921900.

\end{document}